\documentclass[aps,prb,twocolumn,groupedaddress,epsfig,psfig,showpacs]{revtex4-1}

\usepackage{graphicx}
\usepackage{amsmath}

\def\CCI{CeCoIn$_{5}$}

\begin{document}

\title{Band dependent emergence of heavy quasiparticles in \CCI}

\author{A. Koitzsch$^{1}$, T. K. Kim$^{1, 2}$, U. Treske$^{1}$, M. Knupfer$^{1}$, B. B\"uchner$^{1,3}$, M. Richter$^{1}$, I. Opahle$^{1,4}$, R. Follath$^{5}$, E. D. Bauer$^{6}$, and J. L. Sarrao$^{6}$}

\affiliation
{$^{1}$IFW-Dresden, P.O.Box 270116, 01171 Dresden, Germany\\
$^{2}$Diamond Light Source Ltd., Didcot, Oxfordshire, OX11 0DE, United Kingdom\\
$^{3}$Institut f\"ur Festk\"orperphysik, Technische Universit\"at Dresden, 01062 Dresden, Germany\\
$^{4}$ICAMS, Ruhr-Universit\"at Bochum, 44780 Bochum, Germany\\
$^{5}$Helmholtz-Zentrum Berlin, Albert-Einstein-Strasse 15, 12489 Berlin, Germany\\
$^{6}$Los Alamos National Laboratory, Los Alamos, New Mexico 87545, USA}

\date{\today\\}

\begin{abstract}
We investigate the low temperature (T $<$ 2 K) electronic structure of the heavy fermion superconductor \CCI (T$_c$ = 2.3 K) by angle-resolved photoemission spectroscopy (ARPES).  The hybridization between conduction electrons and f-electrons, which ultimately leads to the emergence of heavy quasiparticles responsible for the various unusual properties of such materials, is directly monitored and shown to be strongly band dependent. In particular the most two-dimensional band is found to be the least hybridized one. A simplified multiband version of the Periodic Anderson Model (PAM) is used to describe the data, resulting in semi-quantitative agreement with previous bulk sensitive results from de-Haas-van-Alphen measurements.

\end{abstract}

%\pacs{71.18.+y, 74.25.Jb, 74.72.-h}

\maketitle

The properties of heavy fermion materials sharply deviate from conventional metals, most notably in the occurrence of very large effective masses at low temperatures. This is often accompanied by rich phase diagrams comprised of magnetic order and unconventional superconductivity as well as quantum criticality in which the degree of hybridization between the localized atomic f-states and mobile conduction electron states determines the groundstate. It is widely accepted that at high temperatures the f - electrons are expelled from the Fermi surface and act as incoherent scatterers, while they form hybridized bands with the conduction electrons at low temperatures, facilitating the mass increase and a distinct drop in resistivity. 

The family of heavy fermion compounds Ce$\textit{T}$In$_{5}$ ($\textit{T}$ = Co, Ir, Rh) (115 family) are prototypical heavy fermion superconductors in close proximity to antiferromagnetism, as exemplified by \CCI, which becomes superconducting below T$_c$ = 2.3 K and is located near an antiferromagnetic quantum critical point \cite{Petrovic2001, Bianchi2003, Paglione2003}. 
Phenomenological models indicate that the two-dimensional electronic structure (i.e., cylindrical bands) may be beneficial for the unconventional superconductivity \cite{Mathur1998, Monthoux2001}. On the other hand, a recent proposal based on dynamical mean field theory (DMFT) calculations concluded that the f-hybridization in CeIrIn$_{5}$ is essentially three dimensional \cite{Shim2007}. Thus,  it is not clear that the f - conduction electron hybridization is strongest on the two-dimensional bands or if it is equally strong on all bands close to the Fermi level. The experimental clarification of this point, which is important for the whole class of 115 compounds and for the physical understanding of the heavy fermion superconductors in general, is the main objective of this letter.

To this end we perform angle-resolved photoemission spectroscopy (ARPES) to elucidate the low energy electronic structure of \CCI. Generally, the bandstructure of this compound is reasonably described by LDA (local density approximation) calculations, except for the low energy region, where the influence of the f-electrons becomes important \cite{Fujimori2003, Moore2002, KoitzschCCI_ElSt, Xiao2011, Booth2011}. The latter was studied by ARPES for CeIrIn$_{5}$ and CeCoIn$_{5}$, revealing the existence of hybridized f-bands \cite{Fujimori2006, KoitzschCCIHyb}. However, previous efforts for \CCI\ were restricted to a narrow region in momentum space $\bf{k}$ and to T = 25 K, where the coherent heavy quasiparticle band has not fully developed yet. Here we report on measurements down to T = 1.4 K and study the f-states in a wide k-range. This enables us to quantitatively determine the degree of f-hybridization for each band individually. We find indeed a pronounced band-dependence of the hybridization.
All photoemission measurements were performed at the BESSY 1$^{3}$ ARPES end station equipped with a SCIENTA R4000 analyzer and a Janis $^{3}$He cryostat. Spectra presented in this manuscript were recorded from high quality \CCI\ single crystals cleaved at low temperature. The energy resolution was $\Delta$E = 22 meV at h$\nu$= 121 eV. Single crystals of \CCI\ were grown in In flux \cite{Petrovic2001}. The bandstructure calculation was performed using the scalar-relativistic version of the full potential local orbital minimum basis bandstructure method \cite{Koepernik1999, Note6}. Technical details of the calculations have been described in Ref. \cite{Elgazzar2004}.

\begin{figure*}[t]
\includegraphics[width=0.95\linewidth]{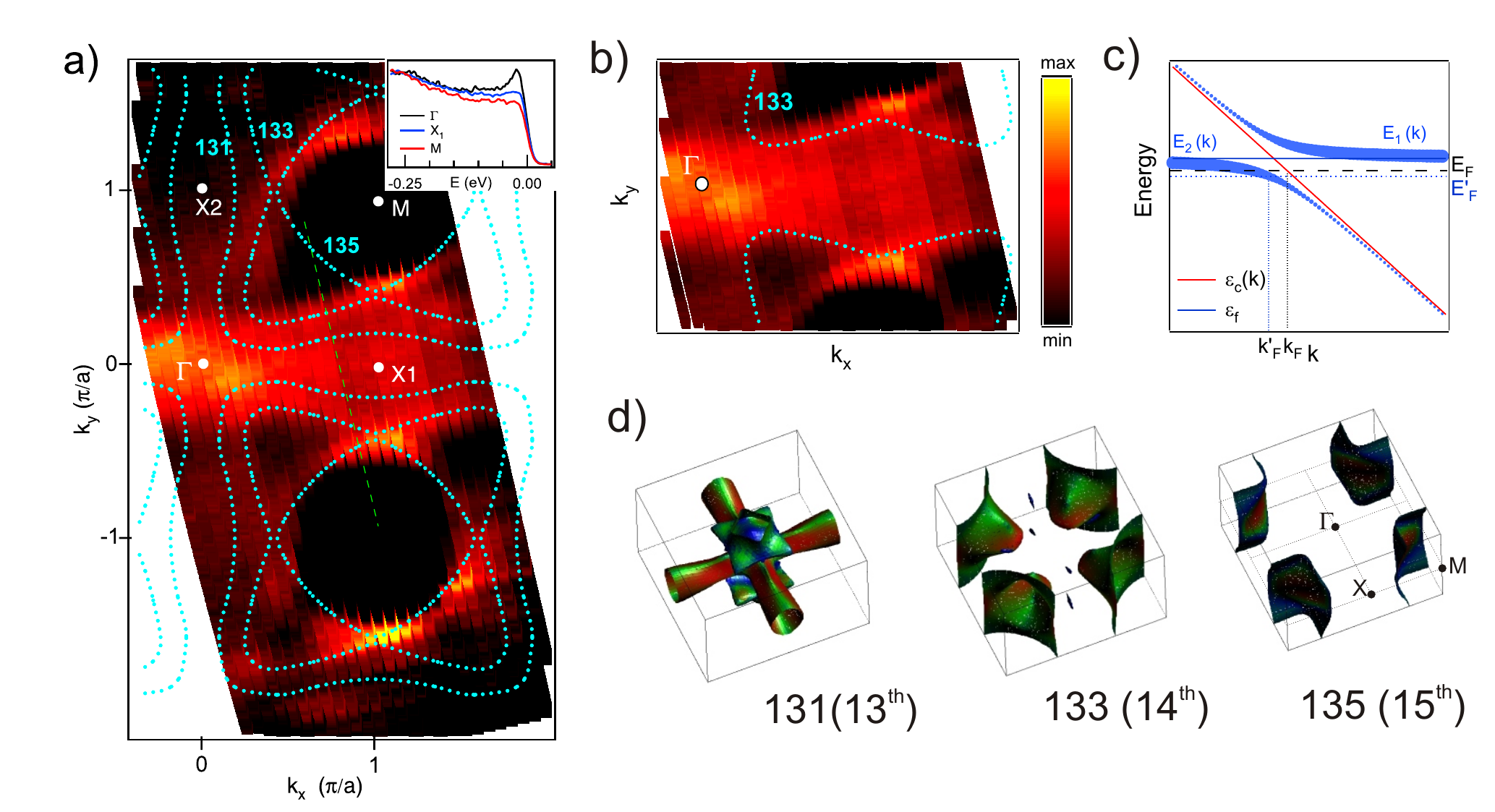}% Here is how to import EPS art
\caption{(Color online) a) Fermi surface map of \CCI\ on resonance (h$\nu$ = 121 eV) at T = 39 K. Blue dashed lines represent LDA calculations without f-contributions. Green dashed line highlights the k-region of the measurement in Fig. 3. (Inset) EDCs from high symmetry points. b) Detail of the Fermi surface. c) Scheme of the f - hybridization with the conduction band. $\varepsilon$$_{c}$(k) is the dispersion of the unhybridized conduction band,  $\varepsilon$$_{f}$ is the energy of the unhybridized f-level, E$_{1,2}$(k) are the hybridized bands. E$_{F}$, E'$_{F}$ and k$_{F}$, k'$_{F}$ are the positions of the Fermi level and Fermi wavevector in the initial and hybridized state. d) Fermi surfaces (LDA without f-contribution). In brackets: Band notation according to Ref. \cite{Settai2001}.} \label{fig:Fig1}
\end{figure*}

Signs of hybridization between f- and conduction electrons are found in several ways in the photoemission data. 
Fig. 1a shows the Fermi surface of \CCI\ measured with a photon energy of h$\nu$ = 121 eV at a temperature of T = 39 K compared to LDA bandstructure calculations, where the f-electrons are treated as localized states and k$_{z}$ $\approx$ (1/8) $\pi$/c. 
At h$\nu$ = 121 eV the cross section of the 4f emission is resonantly enhanced due to the Ce 4d $\rightarrow$ 4f absorption threshold. The LDA bands 133 and 135 fit the ARPES data reasonably well. Band 131 is more difficult to assign with certainty as it is situated in regions with blurred intensity around X$_1$ \cite{Note1}. Fig. 1b shows a magnification of this region. Clearly, band 133 forms a boundary of the blurred intensity around X$_1$. This behavior of band 133 is a sign of the f-hybridization. It can be rationalized with the help of a generic two level mixing model which mimics the periodic Anderson model (PAM) considered an appropriate description of the hybridization between the f-electron and conduction electron states (see Fig. 1c) \cite{Hewson1997}: The flat f-band, located initially above E$_{F}$, mixes with the strongly dispersing conduction band which opens a hybridization gap and redistributes the orbital dependent spectral weight. k$_{F}$ changes so as to accommodate the additional f-electrons in the Fermi surface volume. Most importantly the region of large f-weight below E$_{F}$ (where it can be observed by photoemission) appears mainly at one side of the Fermi surface crossing, namely "outside" of the original conduction band dispersion. 
The distribution of spectral weight in Fig. 1b is indeed confined "outside" of band 133 and thus a signature of sizable hybridization of band 133 with the f-electron system. In contrast, for band 135 this effect is not observed, meaning that the f-hybridization of band 135 is weaker.  

Fig. 1d presents the three dimensional Fermi surfaces of band 131, 133 and 135. Band 135 consists of quasi two dimensional cylinders at the zone corners. Bands 131 and 133 show much stronger k$_{z}$ dispersion. A previous DMFT study of CeIrIn$_5$ suggested that the f-hybridization was strongest in the out-of-plane Ce-In(2) atom-pair compared to the planar Ce-In(1) pair\cite{Shim2007}. This is in agreement with our results. The distribution of f-weight at the Fermi surface is governed by the more three dimensional bands.

Fig. 2 presents the measured intensity as a function of energy along the $\Gamma$M high symmetry direction on-resonant (h$\nu$  = 121 eV, Fig. 2a) and off-resonant (h$\nu$  = 117 eV, Fig. 2c) at T = 1.7 K. The valence bands are again well described by band structure calculations. 
The itinerant treatment of the f-electrons in the calculation decreases the bandwidth slightly with respect to the localized treatment which fits the experiment somewhat better. A clear deviation of LDA from experiment occurs in the region highlighted by the white dashed rectangle in Fig. 2a where the intense flat f-bands occur. By contrast, the flat f-band is absent in the off-resonant spectra presented in Fig. 2c.  Figs. 2 b, d show the momentum integrated intensity related to Figs. 2 a, c. For the on-resonant spectra (Fig. 2b) the typical line shape of hybridized Ce systems is observed: the f$^{1}_{5/2}$ peak at  E$_{F}$ and its spin-orbit split counterpart at about E = - 280 meV. Both features are absent for the off-resonant spectra.
Fig. 2a confirms our findings from the discussion of the Fermi surface: bands 131 and 133 are strongly renormalized near E$_{F}$ while band 135 remains much less hybridized. This is even more clear in the background subtracted version of Fig. 2a in Fig. 3a.

\begin{figure}[h!]
\includegraphics[width=1\linewidth]{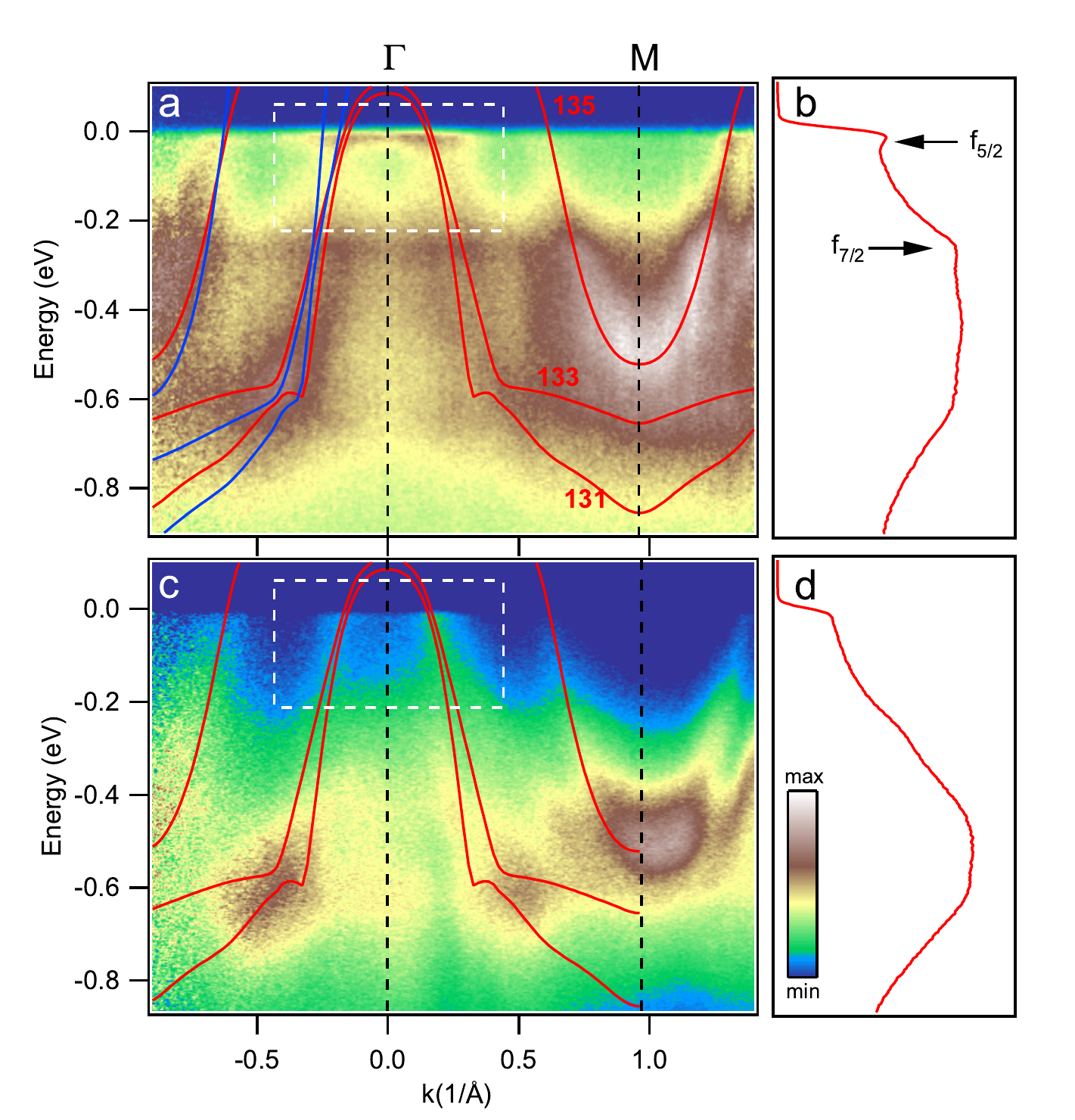}% Here is how to import EPS art
\caption{(Color online) Energy distribution map along $\Gamma$M taken at T = 1.7 K compared to bandstructure calculations on-resonant (a) and off resonant (c). Red lines correspond to LDA calculation treating the f-electrons as being itinerant, blue lines as localized. (b), (d) Momentum integration of (a), (c). Note the renormalized flat bands near E$_{F}$ in (a) highlighted by the dashed white rectangle, which are absent in (c).} \label{fig:Fig2}
\end{figure}

For a quantitative interpretation of the low energy electronic structure of heavy fermion systems as observed by ARPES experiments hybridization models have been employed previously \cite{Im2008, Klein2011,Vyalikh2009,  Danzenbaecher2011}. In our case several bands are involved in the hybridization process. For a first description we may omit band 135, but bands 131 and 133 have to be taken into account simultaneously. Therefore, we extend the two-level mixing model as sketched in Fig. 1c to a three level mixing, which yields the Hamiltonian:

\begin{equation}
H=\begin{pmatrix}
 \varepsilon^{131}_c(k) & 0 & V_{cf}^{131}\\
 0 & \varepsilon^{133}_c(k)& V_{cf}^{133}\\
V_{cf}^{131} & V_{cf}^{133} & \varepsilon_f
\end{pmatrix}
\label{eqn:formel}
\end{equation}

Here  $\varepsilon^{131}_c(k)$ and $\varepsilon^{133}_c(k)$ are the extrapolated experimental dispersion relations of the unhybridized bands \cite{Suppl}. $\varepsilon_f$ is the energy of the f-level and $V_{cf}^{131}$ and $V_{cf}^{133}$ are the hybridization parameters of band 131 and 133 respectively.

\begin{figure*}[t]
\includegraphics[width=0.95\linewidth]{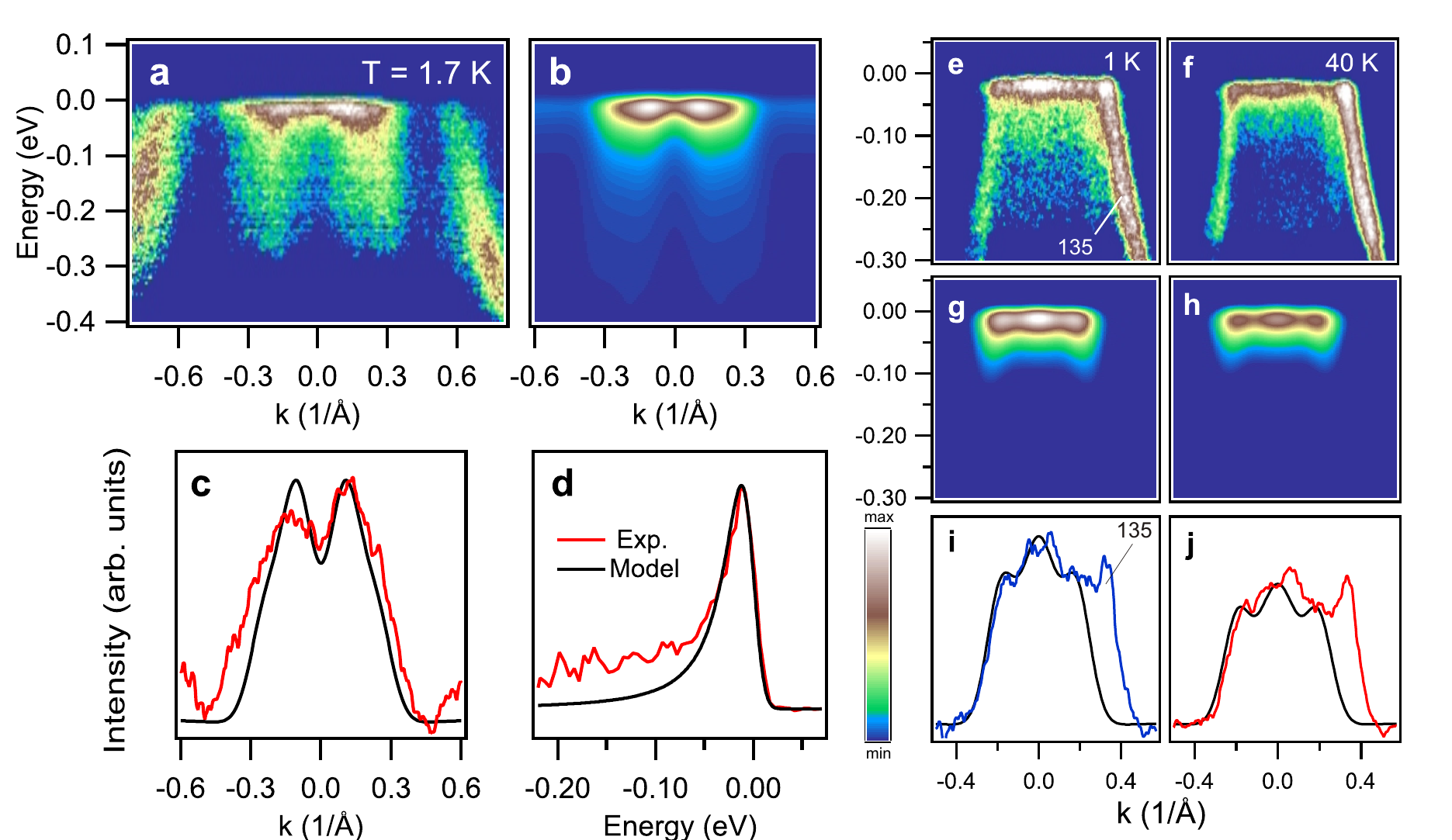}% Here is how to import EPS art
\caption{(Color online) a) Expanded and background subtracted view on the electronic structure near $\Gamma$M. An EDC from the M-point has been taken as background intensity and aligned at higher energies before subtraction. b) Simulation of the f-related spectral function using Eq. (1). c) Comparison of MDCs at E$_F$ and EDCs at k = 0 (d). e, f) Energy distribution maps near MX at T = 1.4 K and 40 K. A background has been subtracted. The bright band at the right hand side is band 135 not considered in the calculations below. (g,h) Simulated spectral function. The same color scale has been used for the pairs (e, f) and (g, h). (i, j) Comparison of experimental and simulated MDCs. } \label{fig:Fig3}
\end{figure*}

Fig 3b shows a solution of this model for optimized parameters $V_{cf}^{131}$ = 50 meV, $V_{cf}^{133}$ = 40 meV and $\varepsilon_f$ = 8 meV. For the details of the calculation and of the simulation of the spectral function see the Supplemental Material \cite{Suppl}. 

Fig. 3c compares the momentum distribution curves (MDC) integrated in a energy window of $\Delta$E = 21 meV below E$_{F}$ and Fig. 3d the energy distribution curves (EDC) at k = 0 ($\Delta$k = 0.004 1/\AA). Note that the high energy tail of the EDC is underestimated by the simulation, as the contribution of the non-f conduction bands is omitted. This is also apparent in Fig. 3a in form of the non-vanishing intensity between E = -0.1 eV and -0.25 eV, absent in Fig. 3b. It is interesting that the width of the EDC peak is not resolution limited. Additional phonon broadening has been proposed previously for other Ce compounds \cite{Arko1999}, but the crystal field splitting could play also a role in the present case. The energy separation of the crystal field levels are 6.8 meV and 25 meV above the ground state \cite{Willers2010}, which is approximately in the order of the applied Lorentzian broadening (40 meV). 

In Fig. 3 e,f cuts of the electronic structure near the MX direction (see green dashed line in Fig. 1) for different temperatures are presented. Again flat f-derived intensity is found in-between the bands crossing E$_F$. The situation is complicated in this k-region by the fact, that band 135 (the bright steeply dispersing object on the right hand side) is close to bands 131 and 133. Nevertheless, the spectral function has a characteristic form and a clear temperature dependence. Figs. 3 g,h show the results of the model calculations with hybridization parameters ($V_{cf}^{131}$; $V_{cf}^{133}$) = (30 meV; 50 meV) at T =1.4 K and ($V_{cf}^{131}$; $V_{cf}^{133}$) = (29 meV; 42 meV) at 40 K. 
The decrease of the hybridization parameters with increasing temperature is in agreement with the expectation that the hybridization vanishes at high temperatures. 

Experimental and simulated MDCs in the same energy window as used in Fig. 3c are compared in Fig. 3 i, j. The simulation overestimates somewhat the momentum broadening of band 131 as is seen from the observed splitting of the inner peak. Note, that the datasets in Figs. 3e and f, from which the MDCs have been extracted, are normalized at higher energies to each other. This intensity ratio has been preserved in the presentation of the MDCs in Figs. 3 i, j. 

The presented information on the electronic structure and the extracted k and temperature dependent hybridization parameters must be reflected in a wide variety of macroscopic properties of the solid. Table I compares our results to previous de-Haas-van-Alphen (dHvA) measurements \cite{Settai2001, Hall2001}. The Fermi surface areas measured from Fig. 1 match the dHvA frequencies, establishing the bulk nature of our measurements and confirming our previous results for the top-plane of the Brillouin zone \cite{KoitzschCCI_ElSt}. 
The mass enhancement has been calculated from the ratio of the renormalized and unrenormalized Fermi velocities v$_{F}$ for $\Gamma$M and MX separately. The renormalized Fermi velocities were obtained from the simulation, the unrenormalized Fermi velocities from extrapolation of the conduction bands (see Supplemental Material for further description and a discussion of the absolute values of m* as derived from the model) \cite{Suppl}. Taken at face value the mass enhancements match approximately the dHvA result for the $\beta_{1}$ orbit, which is a momentum average and has been evaluated from the ratio of the measured cyclotron mass and the band-mass m$_{b}$ \cite{Settai2001}. However, the cyclotron mass shows a field dependence and is expected to be larger in the zero field limit probed here. Nevertheless, the agreement is satisfactory for band 133. But the dHvA measurements reveal also a significant mass enhancement of band 135 , which we have neglected so far, based on account that the hybridization is weaker than for bands 133 and 131.

 \begin{table*}
 \caption{\label{tab:Table2} Physical quantities derived from ARPES and dHvA. The Fermi surface area (converted to dHvA frequency units) has been evaluated from Fig. 1. The mass enhancement factors for $\Gamma$M and MX directions are results from our simulations.}
 \begin{ruledtabular}
 \begin{tabular}{ccccccccc}
& &\multicolumn{3}{c}{Frequency (kT)}&\multicolumn{4}{c}{m*/m$_{b}$}\\
band & Orbit & Ref.\cite{Settai2001}  & Ref.\cite{Hall2001} & present & Ref.\cite{Settai2001}  & $\Gamma$M &  MX \\
\hline
133 & $\beta$$_{1}$ & 12.0 & - & 10.5-12 & 13.8  & 8.2 & 14.7  \\
135 & $\alpha$$_{2}$ & 4.53 & 5.16 & 4.2-4.7 & 16.5 & - & - \\
 \end{tabular}
 \end{ruledtabular}
 \end{table*}

The opening of the hybridization gap at low temperature has been directly observed by optical spectroscopy \cite{Singley2002, Mena2005, Burch2007} and tunneling spectroscopy \cite{Aynajian2012}. The description of the low temperature optical conductivity of \CCI\ requires a broad spectrum of hybridization gaps \cite{Burch2007}, ranging from $\approx$ 50 cm$^{-1}$ (6 meV) to 400 cm$^{-1}$ (48 meV). This can be directly compared to our V$_{cf}$ values. We notice that the low energy region is not covered by our results. Together with the reported mass enhancement of the $\alpha$ orbits of band 135 we suspect that band 135 has a small hybridization gap, which is below our detection limit. 
Quasiparticle interference patterns from tunneling spectroscopy have been interpreted in terms of the hybridization gap opening of band 135 but in a different k$_z$ plane than considered here \cite{Aynajian2012}. It is indeed possible that the hybridization has a k$_z$ dependence. This has been incorporated in a previous theoretical investigation of the superconducting properties of \CCI\ using a 3D PAM for band 133 \cite{Tanaka2006}. Band 133 formed also the basis for the RPA (random phase approximation) description of a sharp resonance mode observed by neutron scattering \cite{Stock2008}. This mode is intimately related to superconductivity \cite{Eremin2008}. Thus, important aspects of superconductivity are related to three dimensional features of the electronic structure. On the other hand it is known that magnetic fluctuations decay more slowly in two dimensions, in agreement with the quasi two-dimensional properties of \CCI. If the hybridization of the 2D electrons is weaker, the Kondo screening of the magnetic moments will be weaker, hence allowing for increased 2D magnetic fluctuations. 

The present study demonstrates that a consistent description of the low energy electronic structure of a model heavy fermion system can be achieved by relatively simple assumptions based on the periodic Anderson model. The f-hybridization in \CCI\ as measured by ARPES is quantified and appears to be band and k-dependent, in agreement with previous theoretical proposals. We find that the most pronounced two dimensional band shows the weakest f-hybridization, suggesting that further research is necessary to reveal the delicate interplay between superconductivity, magnetism and the underlying electronic structure in heavy fermion materials.

A.K. acknowledges financial support by the Deutsche Forschungsgemeinschaft (Grant No. KO 3831/1-1).
Work at Los Alamos was performed under the auspices of the U.S. Department of Energy, Office of Science, Division of Materials Science and Engineering. A.K. thanks S. V. Borisenko, A. Akbari and P. Thalmeier for discussions.

\bibliography{pap_condmat}

\end{document}